\documentclass[english,aps,nofootinbib,twocolumn,preprintnumbers,nobalancelastpage,floatfix,groupedaddress,superscriptaddress,prl,reprint]{revtex4-1}

\usepackage[latin1]{inputenc}
\pdfoutput = 1
\usepackage{bm}
\usepackage{color}
\usepackage{graphicx}
\usepackage{cancel}
\usepackage{amsfonts}
\usepackage{amssymb}
\usepackage{amsmath}
\usepackage{calc}
\usepackage{dcolumn}
\usepackage{mathrsfs}

\usepackage[normalem]{ulem}             

\newcommand{\virg}[1]{`#1'}
\newcommand{\wirg}[1]{``#1''}
\newcommand{\eg}{e.g.~}
\newcommand{\ie}{i.e.~}

\newcommand{\Eq}[1]{Eq.~\eqref{#1}}

\newcommand{\Lag}{\mathscr{L}}	

\newcommand{\beq}{\begin{equation}}
\newcommand{\eeq}{\end{equation}}
\newcommand{\ud}{\text{d}}

\newcommand{\ER}{E_\text{R}}
\newcommand{\mDM}{m_\text{DM}}
\newcommand{\gDM}{g_\text{DM}}
\newcommand{\vesc}{v_\text{esc}}
\newcommand{\vmin}{v_\text{min}}

\definecolor{rossoCP3}{cmyk}{0,.88,.77,.40}
\definecolor{verdeCP3}{rgb}{0.09765625, 0.57421875, 0.1015625}
\definecolor{bluCP3}{rgb}{0, 0.23, 0.67}

\ifx\pdfoutput\undefined
\usepackage[dvips,bookmarks]{hyperref}    
\else
\usepackage{hyperref}    
\fi
\hypersetup{colorlinks, bookmarksopen, bookmarksnumbered,
citecolor=verdeCP3, linkcolor=bluCP3, pdfstartview=FitH, urlcolor=rossoCP3}

\newcommand{\AddrIAP}{%
Institut d'Astrophysique de Paris, 98bis boulevard Arago, 75014 Paris (France)}
\newcommand{\AddrGRAPPA}{%
GRAPPA Institute, University of Amsterdam, Science Park 904, 1090 GL Amsterdam (Netherlands)}
\newcommand{\AddrUCLA}{%
Department of Physics and Astronomy, UCLA, 475 Portola Plaza, Los Angeles, CA 90095 (USA)}

\begin{document}

\title{Dark matter with pseudo-scalar-mediated interactions explains \\ the DAMA signal and the Galactic Center excess}

\author{Chiara~Arina}
\affiliation{\AddrGRAPPA}
\affiliation{\AddrIAP}
\author{Eugenio~Del~Nobile}
\affiliation{\AddrUCLA}
\author{Paolo~Panci}
\affiliation{\AddrIAP}

\begin{abstract}
We study a Dirac Dark Matter particle interacting with ordinary matter via the exchange of a light pseudo-scalar, and analyze its impact on both direct and indirect detection experiments. We show that this candidate can accommodate the long-standing DAMA modulated signal and yet be compatible with all exclusion limits at $99_S$\% CL. This result holds for natural choices of the pseudo-scalar-quark couplings (e.g.~flavor-universal), which give rise to a significant enhancement of the Dark Matter-proton coupling with respect to the coupling to neutrons. We also find that this candidate can accommodate  the observed $1$\,--\,$3$ GeV gamma-ray excess at the Galactic Center and at the same time have the correct relic density today. The model could be tested with measurements of rare meson decays, flavor changing processes, and searches for axion-like particles with mass in the MeV range.
 \end{abstract}

\maketitle

\section{Introduction}\label{sec:intro}

Direct Dark Matter (DM) search experiments have underwent astonishing developments in recent years, achieving unprecedented sensitivity to Weakly Interacting Massive Particles (WIMPs) in the mass range from few GeV to tens of TeV. The most stringent limits on the DM parameter space are set by LUX~\cite{Akerib:2013tjd}, XENON100~\cite{Aprile:2012nq}, and SuperCDMS \cite{Agnese:2014aze} for spin-independent interactions, with PICASSO~\cite{Archambault:2012pm}, SIMPLE~\cite{Felizardo:2011uw}, COUPP~\cite{Behnke:2012ys}, and KIMS \cite{Kim:2012rza} setting relevant bounds for spin-dependent interactions and DM-proton couplings. While these and other searches did not find evidences for DM, four experiments have signals that can be interpreted as due to WIMP scatterings \cite{Aalseth:2014eft, Aalseth:2014jpa, Angloher:2011uu, Agnese:2013rvf}. The significance of the excesses is mild (from $2\sigma$ to $4\sigma$), except for DAMA's result \cite{Bernabei:2013xsa}, where the observation of an annually modulated rate as expected from the simplest model of DM halo, reaches the very high significance of $9.3 \sigma$. This achievement however has received a long-standing series of criticisms, given that the interpretation of the DAMA data in the light of many models of WIMP interactions is incompatible with all exclusion bounds.

Another  claim of possible evidence of WIMP interactions comes from a $1$\,--\,$3$ GeV $\gamma$-ray excess observed in the Galactic Center (GC)~\cite{Daylan:2014rsa} by the Fermi satellite. Although milli-second pulsars may be responsible for explaining the excess~\cite{Abazajian:2014fta}, the possibility of DM annihilation has attracted a lot of attention by the community. In fact, the excess can be fitted with models of annihilating DM which roughly provide the correct thermal relic density.
In \cite{Boehm:2014hva} for instance it was shown that a Dirac WIMP interacting with Standard Model (SM) fermions through a pseudo-scalar mediator can achieve the desired annihilation cross section, avoiding at the same time constraints from DM collider searches, cosmic antiprotons and solar neutrino fluxes, and the cosmic microwave background. In fact, the point of Ref.~\cite{Boehm:2014hva} is that the DM might be \virg{Coy}, meaning that it can have a single detectable signature (in this case the annihilation into $\gamma$-rays) while escaping all other searches.

In this letter we show that Coy DM with a light pseudo-scalar mediator can fit at the same time the GC $\gamma$-ray excess and the DAMA data, while being compatible with all null direct detection experiments.

\section{The Dark Matter model}\label{sec:mod}

The DM is a Dirac fermion $\chi$ with mass $\mDM$, which interacts, with a coupling $g_{\rm DM}$, with a (real) pseudo-scalar $a$ with mass $m_a$ coupled to the SM fermions:
\beq\label{eq:lagrangian}
\Lag_\text{int} = -i \frac{g_{\rm DM}}{\sqrt{2}} a \bar{\chi}\gamma_5\chi - i g \sum_f \frac{g_f}{\sqrt{2}} a \bar{f} \gamma_5 f \,.
\eeq
In the following we will consider two types of fermion couplings $g_f$: flavor-universal couplings $g_f = 1$ independent of the fermion type, 
and Higgs-like couplings proportional to the fermion masses $g_f =  m_f / 174$ GeV. Furthermore, for the direct detection analysis we will consider also the case of DM coupled equally to protons and neutrons (isoscalar interaction,\footnote{Notice that our use of the term \virg{isoscalar} refers to the isospin symmetry between proton and neutron. As it will become clear later on this does not imply, nor is implied by, isospin symmetry at the quark level.} also called \wirg{isospin-conserving}), as assumed \eg by~\cite{Gresham:2014vja,Catena:2014uqa}. In all cases we denote with $g$ a multiplicative factor common to all couplings of $a$ with SM fermions.

\section{Direct detection}\label{sec:dd}
When computing scattering cross sections at direct detection experiments, it is necessary to bear in mind that the scattering occurs with the whole nucleus due to the small WIMP speed. Therefore, starting with an interaction Lagrangian with quarks as in \eqref{eq:lagrangian}, one needs first to determine the DM-nucleon effective Lagrangian and then to properly take into account the composite structure of the nucleus which results in the appearance of nuclear form factors in the cross section.

The first step is accomplished in our case by taking the following effective DM-nucleon interaction Lagrangian, valid in the regime of contact-interaction:
\beq\label{eq:lagrangianN}
\Lag_\text{eff} = \frac1{2\Lambda_a^2}\sum_{N = p, n} g_N \, \bar\chi \gamma^5 \chi \, \bar N \gamma^5 N \,, 
\eeq
where $\Lambda_a \equiv m_a / \sqrt{g_{\rm DM} g}$.  The proton and neutron coupling constants are given by
\beq\label{eq:sumq}
g_N = \sum_{q = u, d, s} \frac{m_N}{m_q} \biggl[ g_q - \sum_{q' = u, \dots, t} g_{q'} \frac{\bar{m}}{m_{q'}} \biggr] \Delta_q^{(N)} \,,
\eeq
where $\bar{m} \equiv (1 / m_u + 1 / m_d + 1 / m_s)^{-1}$ and we use
\beq\label{eq:Delta}
\begin{split}
\Delta_u^{(p)} &= \Delta_d^{(n)} = + 0.84 \,,
\\
\Delta_d^{(p)} &= \Delta_u^{(n)} = - 0.44 \,,
\\
\Delta_s^{(p)} &= \Delta_s^{(n)} = - 0.03
\end{split}
\eeq
for the quark spin content of the nucleon~\cite{Cheng:2012qr}. 

It is important to notice here that $g_p$ is {\it naturally} larger (in modulus) than $g_n$ in both the flavor-universal and Higgs-like coupling scenarios. This will have important phenomenological consequences. In fact, since the interaction \eqref{eq:lagrangianN} measures a certain component of the spin content of the nucleus carried by nucleons \cite{Fitzpatrick:2012ix}, a large $g_p / g_n$ will favor those nuclides (like $^{23}$Na, $^{127}$I and $^{19}$F) with a large spin due to their unpaired proton rather than $^{129, 131}$Xe nuclei with an unpaired neutron. Given that the most stringent bounds for most DM-nucleus interactions are given at present by experiments using xenon (LUX, XENON100)\footnote{We do not consider germanium detectors as their sensitivity to spin-dependent interaction via unpaired protons is smaller than \eg COUPP in the mass range relevant for Coy DM.} while DAMA employs sodium and iodine, a large value of $g_p / g_n$ would go in the direction of reconciling them. From the values in \eqref{eq:Delta} we get $g_p / g_n = -16.4$ for flavor-universal and $-4.1$ for Higgs-like interactions. The relative size of the two couplings depends on the actual values of the $\Delta_q^{(N)}$'s, which are uncertain (see \eg Table 4 in~\cite{DelNobile:2013sia} for a comparison of the different values found in the literature); the values in \eqref{eq:Delta} are conservative in the sense that they minimize the ratio $g_p / g_n$, respect to what obtained with other choices of the $\Delta_q^{(N)}$'s (a second set of values from \cite{Cheng:2012qr}, which brackets from above the possible values of $g_p / g_n$, yields a coupling ratio which is 2.7 and 1.3 times larger than the one given by \eqref{eq:Delta}, for flavor-universal and Higgs-like couplings respectively). Notice that, as long as $g_u = g_d = g_s$, the contribution of the light quarks cancels in~\eqref{eq:sumq}, and one may therefore set $g_u = g_d = g_s = 0$ as in hadronic axion models \cite{Agashe:2014kda}. Finally  we will also use isoscalar interactions, \ie by setting $g = g_p = g_n$ without using~\Eq{eq:sumq}, as assumed in~\cite{Gresham:2014vja,Catena:2014uqa}.

\begin{figure*}[t!]
\includegraphics[width=0.32\textwidth,trim=2mm 2mm 3mm 2mm, clip]{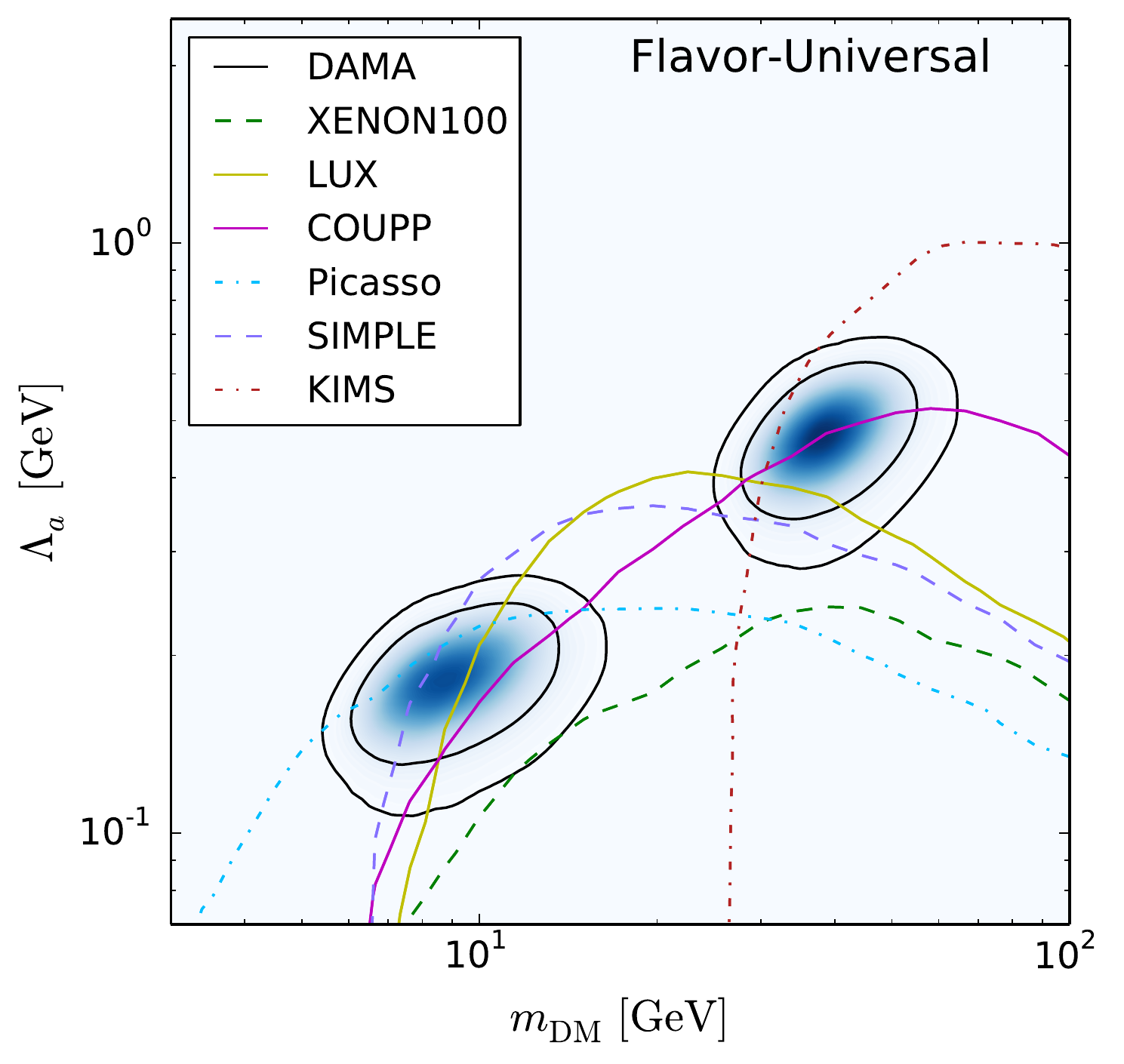} \
\includegraphics[width=0.32\textwidth,trim=2mm 2mm 3mm 2mm, clip]{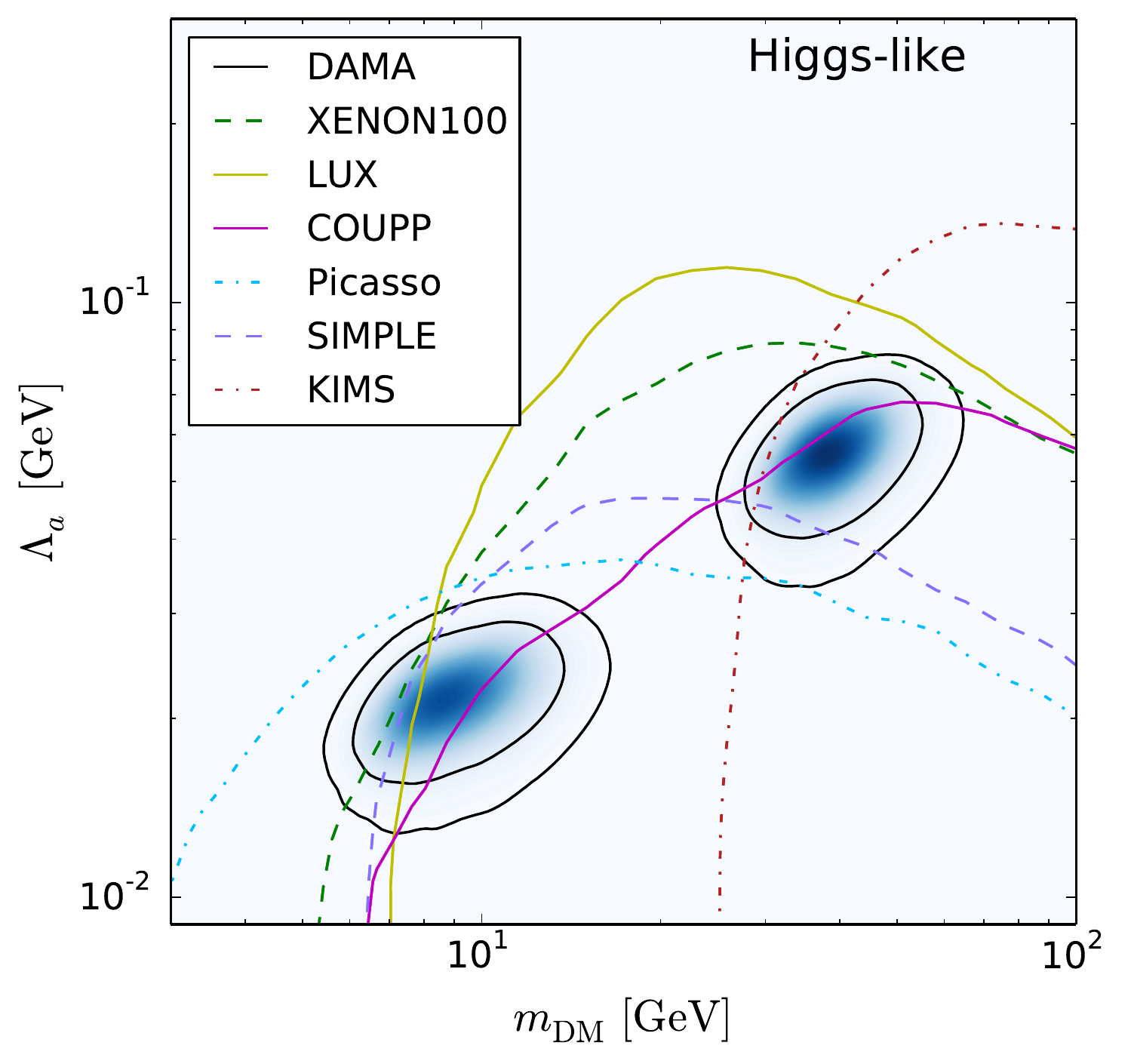} \
\includegraphics[width=0.32\textwidth,trim=2mm 2mm 3mm 2mm, clip]{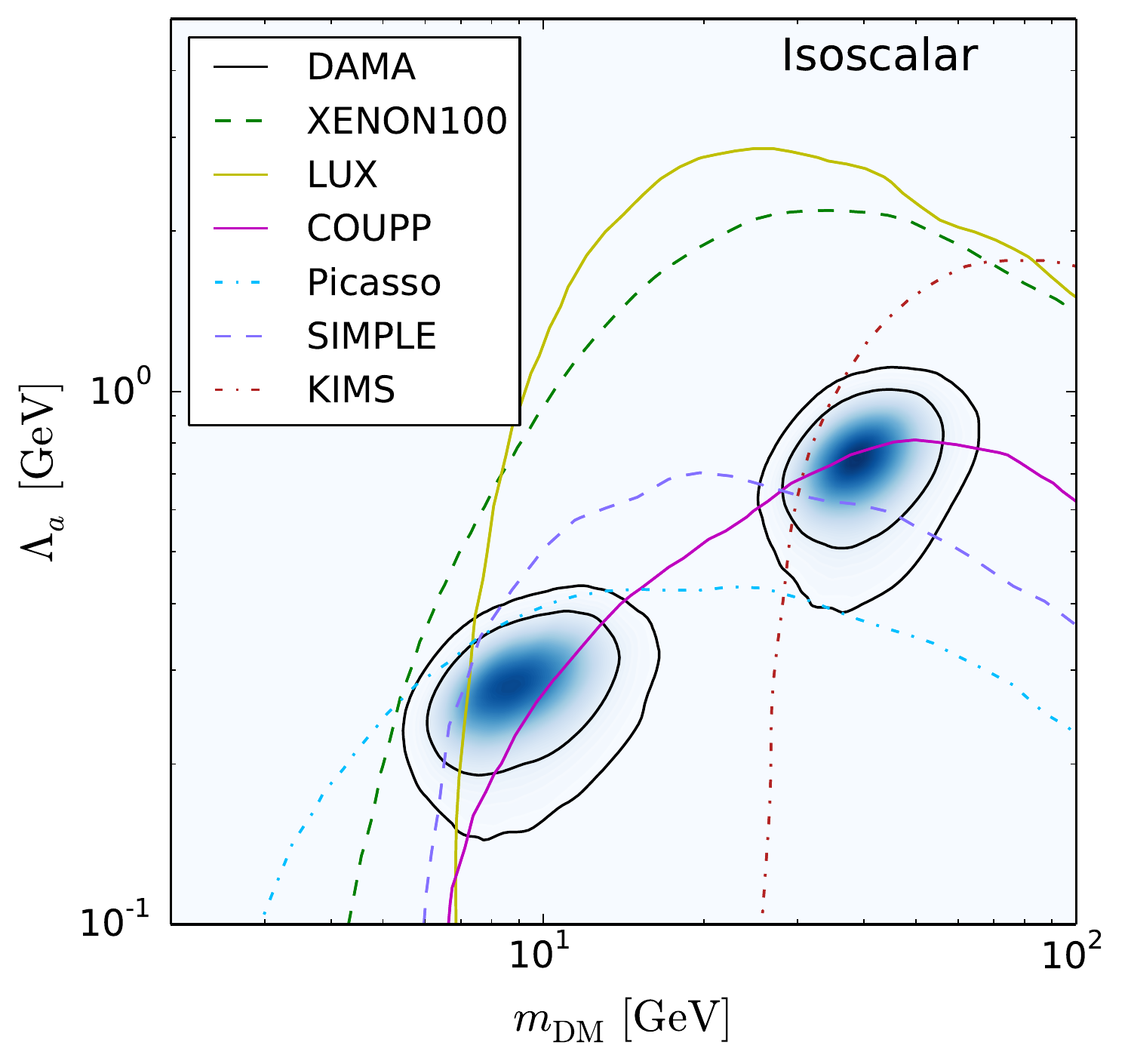}
\caption{\label{fig:FU} 2-dimensional credible regions for DAMA (shaded/black solid, 90\% and 99\% CL) and exclusion limits ($99_S\%$ CL) in the $( \mDM, \Lambda_a )$ plane, for flavor-universal (left), Higgs-like (center) and isoscalar (right) couplings.}
\end{figure*}

Once the DM-nucleon Lagrangian is established, one needs to determine the DM interaction cross section with the nucleus. This is customarily done by coherently adding the amplitudes of interaction with the different nucleons in the nucleus, and multiplying by an appropriate nuclear form factor that parametrizes the loss of coherence in the scattering with increasing exchanged momentum. While form factors for the standard spin-independent and spin-dependent interactions have been extensively studied, little is known of form factors for other interactions. Notice that the Lagrangian \eqref{eq:lagrangianN} corresponds in the non-relativistic limit to a DM-nucleon interaction $(\vec{S}_\chi \cdot \vec{q}) (\vec{S}_N \cdot \vec{q})$, with $\vec{S}_\chi$, $\vec{S}_N$ and $\vec{q}$ the DM spin, nucleon spin and exchanged momentum respectively, while the standard spin-dependent interaction corresponds to $\vec{S}_\chi \cdot \vec{S}_N$. At the nuclear level, the difference stands in the fact that the former interaction only measures the component of the nucleon spin in the nucleus that is longitudinal to $\vec{q}$, while the latter couples to both longitudinal and transverse components. Therefore it is not justified to use the standard spin-dependent form factor for the interaction in \eqref{eq:lagrangianN} as done \eg in \cite{Boehm:2014hva, Chang:2009yt}, although in some cases it could be used as a proxy \cite{Gresham:2014vja}. The form factor to be used in this case has been computed in \cite{Fitzpatrick:2012ix} using standard shell model techniques.

The DM interaction cross section with a target nucleus with mass $m_T$ is
\beq\label{eq:dsder}
\frac{\ud \sigma_T}{\ud \ER} = \frac{1}{128 \pi} \frac{q^4}{\Lambda_a^4} \frac{m_T}{\mDM^2 m_N^2} \frac{1}{v^2} \hspace{-1mm} \sum_{N, N' = p, n} \hspace{-2mm} g_N g_{N'} F_{\Sigma''}^{(N, N')}(q^2)
\,,
\eeq
with $v$ the DM speed in Earth's frame, $\ER = q^2 / 2 m_T$ the nuclear recoil energy and $F_{\Sigma''}^{(N,N')}$ the (squared) form factors. The large suppression factor $q^4/m_a^4$ for large mediator mass is the reason why the interaction in \eqref{eq:lagrangianN} has often been neglected. Given this suppression in the non-relativistic limit, one should check that radiative corrections do not produce unsuppressed interactions that are therefore comparable to the Born cross section at low velocities; however the Lagrangian~\eqref{eq:lagrangianN} is known to not produce such interactions~\cite{Freytsis:2010ne}. It should also be checked that higher order QCD corrections do not spoil the enhancement of the WIMP-proton coupling respect to the WIMP-neutron one, as from Eq.~\eqref{eq:sumq} which is valid at lowest order \cite{Cirigliano:2012pq,Cirigliano:2013zta}. However, since pseudo-scalar currents can only be coupled to an odd number of mesons as opposed \eg to scalar currents, we only expect potential $10 \%$ corrections \cite{CiriglianoPC}.

The scattering rate is
\beq\label{eq:Rate}
\frac{\ud R_T}{\ud \ER} = \frac{\xi_T}{m_T} \frac{\rho}{\mDM} \int_{v \geqslant \vmin} \hspace{-.40cm} \ud^3 v \, v \, f(\vec v) \frac{\ud \sigma_T}{\ud \ER} \,,
\eeq
with $\xi_T$ the target's mass fraction in the detector, $\rho$ the local DM density, and $f(\vec v)$ the DM velocity distribution in Earth's frame, corresponding to a truncated Maxwell-Boltzmann with characteristic speed $v_0 $ and escape velocity $\vesc$ in the galactic frame. Considering elastic scattering and denoting with $\mu_T$ the DM-nucleus reduced mass, $\vmin = \sqrt{m_T \ER / 2 \mu_T^2}$ is the minimum speed a WIMP needs in order to impart the target nucleus with a recoil energy $\ER$. In order to compare with the experimental results, the rate in \eqref{eq:Rate} must be convolved with the detector resolution function and the experimental efficiency (see \eg~\cite{DelNobile:2013sia,Panci:2014gga}).

We analyze data by LUX, XENON100, PICASSO, SIMPLE, COUPP, KIMS and DAMA. We use Bayesian statistics to infer the $99_S\%$ credible interval for the exclusion limits and both the 90\% and 99\% credible regions for DAMA from the posterior probability density function, as detailed in~\cite{Arina:2011si,Arina:2012dr} where it was demonstrated that the procedure is robust against the choice of prior and matches well a profile likelihood analysis. We consider log priors for both our relevant parameters: the DM mass $m_{\rm DM}$, from 1 GeV to 1 TeV, and the scale $\Lambda_a$, from 0.01 GeV to 100 GeV, not to favor a particular mass scale range. For each experiment we marginalize over the nuisance parameters, given by the uncertain astrophysical parameters $\rho$, $v_0$, $\vesc$ (the central values for the Gaussian priors are $\bar\rho = 0.3 \, \rm GeV/cm^3$, $\bar{v}_0 = 230\,  \rm km/s$ and $\bar{v}_{\rm esc} = 544 \, \rm km/s$), as well as the experimental uncertainties as described in~\cite{Arina:2011si,Arina:2012dr}. The details on the likelihood functions for the LUX and COUPP experiments are provided in the appendix.

Fig.~\ref{fig:FU} shows the results of our analysis for our three choices of couplings: flavor-universal, Higgs-like and isoscalar. The two DAMA regions correspond respectively to scattering off Na (peaked around $\mDM \sim 8$ GeV) and I (peaked around $\mDM \sim 40$ GeV). Part of the regions is compatible with all null experiments for flavor-universal couplings at $99_S$\% CL. Notice how the large enhancement of the WIMP-proton coupling with respect to the WIMP-neutron coupling suppresses the LUX and XENON100 bounds but not COUPP, PICASSO, SIMPLE and KIMS. For Higgs-like couplings the LUX and XENON100 bounds are less suppressed due to the reduced $g_p / g_n$ enhancement, and the exclusion limits disfavor both sodium and iodine regions. In the isoscalar case instead there is no enhancement and DAMA is largely disfavored at $99_S\%$ CL by both XENON100 and LUX.

It is intriguing that the allowed DAMA iodine region lies in the ballpark of DM masses that can account for the $\gamma$-ray GC excess. In the following we investigate whether the two signals can be both accommodated within the Coy DM scenario.

\section{The GC excess}\label{sec:gc}
Various authors reported evidence for an excess of $1$\,--\,$3$ GeV $\gamma$-rays from the GC. Taking as a reference Fig.~15 of~\cite{Daylan:2014rsa}, DM particles with a mass $\mDM \sim 20$\,--\,$40$ GeV annihilating mostly into quarks with a cross section $\langle \sigma v \rangle \sim 1$\,--\,$2 \times 10^{-26}$ cm$^3$/s are shown to fit the spectrum of the observed excess. In particular, the results of the fit are shown for models with flavor-universal and Higgs-like couplings (right panel), and can be then directly compared with our results.\footnote{\label{factor2}Notice that Ref.~\cite{Daylan:2014rsa} assumes, in the definition of the $\gamma$-ray flux, that the DM is self-conjugated. 
This  implies that, in order to predict the same signal in the GC, our cross section needs to be a factor of 2 larger than the one found in Ref.~\cite{Daylan:2014rsa}.}  

In this section we show that the Coy DM interpretation of the DAMA data is compatible with a DM explanation of the GC excess. In fact, $\chi$ can annihilate to SM fermions through $s$-channel pseudo-scalar exchange, thus generating a secondary photon flux. The requirement of fitting the $\gamma$-ray excess can then be used to disentangle the pseudo-scalar mass $m_a$ from the product $g_\text{DM} g$ in $\Lambda_a$, that is the parameter constrained by DAMA. As we will see, there is room in the parameter space favored by DAMA (and allowed by the other experiments) to explain the GC excess, for pseudo-scalar masses $m_a \ll \mDM$. This opens up the possibility to also break the degeneracy between $g_{\rm DM}$ and $g$ by demanding that the correct relic density is achieved in the early universe via $\bar\chi \chi \to \bar f f$ and $\bar\chi \chi \to a a$ annihilations (the latter process being p-wave suppressed today), since the two cross sections have different dependence on $g_{\rm DM}$ and $g$.
 
In summary, from the three observables: ({\it i}\,) DAMA signal in direct searches, ({\it ii}\,) $\gamma$-ray excess in the GC, and ({\it iii}\,) correct relic density obtained by solving the Boltzmann equation, we can fully determine the free parameters of the Coy DM Lagrangian for our choices of pseudo-scalar coupling to SM fermions, flavor-universal and Higgs-like. Formulas for the annihilation cross sections are provided in the appendix. For ({\it ii}\,), unlike direct DM searches, indirect detection signals are different if the DM particles couple democratically with all quarks or just with the heavy ones, and we study these two cases separately. We dub these two scenarios \virg{Universal (\emph{democratic})} and \virg{Universal (\emph{heavy-flavors})}, respectively. We neglect annihilation to leptons as the produced $\gamma$-ray flux is smaller than the one due to annihilation into quarks, at equal couplings; the reduction factor can vary between $2$ and $17$ depending on the choice of the couplings. Notice that coupling to leptons is unessential for the purposes of fitting the GC excess and of studying direct detection experiments, unless much larger than the coupling to quarks. However, leptonic couplings are tightly bound by precision measurements of the electron and muon anomalous magnetic moments. For a pseudo-scalar that only couples to heavy quarks, our model is compatible with these measurements as shown in the appendix.

\begin{table}
\centering
\begin{tabular}{l|c|c}
& $\mDM^{\rm best}$ & $\langle \sigma v \rangle_{\rm best}$  \\
\hline
Universal (\emph{democratic}) & 22 GeV & $1.1 \times 10^{-26}\, \rm cm^3/s$ \\
\hline
Universal (\emph{heavy-flavors}) & 31 GeV & $1.4 \times 10^{-26}\, \rm cm^3/s$ \\
\hline
Higgs-like & 33 GeV & $1.6 \times 10^{-26}\, \rm cm^3/s$ \\
\end{tabular}
\caption{\label{tab:fit} Approximate best fit values of the DM mass and the thermally averaged annihilation cross section extracted from Fig.~15 of Ref.~\cite{Daylan:2014rsa}, for different choices of the pseudo-scalar coupling to SM fermions. The values for the Universal ({\em heavy-flavors}) case have been determined by taking the average of the best fit values for the $b \bar b$ and $c \bar c$ channels.}
\end{table}

Table \ref{tab:fit} reports the approximate best fit values of the DM mass and the thermally averaged annihilation cross section, as extracted from Fig.~15 of Ref.~\cite{Daylan:2014rsa}, for our different choices of $g_f$. Adopting these values, from conditions ({\it i}\,), ({\it ii}\,) and ({\it iii}\,) we obtain the following sets of values of the couplings $g_\text{DM}$ and $g$, together with the corresponding value of $m_a$ from the DAMA iodine best fit point:
\begin{itemize}
\item Universal (\emph{democratic}): $g g_f \simeq 7.7 \times 10^{-3}$, $g_{\rm DM} \simeq 0.64$, and $m_a\simeq 35$ MeV. This scenario is favored by direct detection (see Fig.~\ref{fig:FU}, left), however the DM mass required for the GC excess is outside of the 99\% CL of DAMA iodine region (see Table~\ref{tab:fit}).
\item Universal (\emph{heavy-flavors}):  $g g_f \simeq 1.8 \times 10^{-2}$ for the heavy flavors and $0$ otherwise, $g_{\rm DM} \simeq 0.72$, and $m_a\simeq 56$ MeV. This is the best-case scenario, as the DM mass required to fit the $\gamma$-ray excess is fully compatible with the DAMA iodine signal.
\item Higgs-like: $g g_f \simeq 1.15 \, m_f / 174$ GeV, $g_{\rm DM} \simeq 0.69$, and $m_a \simeq 52$ MeV. Here the GC signal is compatible with the DAMA iodine allowed region, which is however excluded at $99_S$\% CL by LUX and XENON100 as shown in Fig.~\ref{fig:FU} (center).
\end{itemize}

For direct detection, the favored values of the pseudo-scalar mass are of the same order as the typical momentum transfer. Therefore we expect small changes in our fit to DAMA data due to the onset of the long-range regime, however this will not modify our conclusions. Such a light mediator might be problematic in what it could be stable or have a long lifetime (on cosmological time-scales), thus constituting a sizable component of the DM or otherwise injecting unwanted energy after the time of big bang nucleosynthesis. However, the pseudo-scalar state always decays before the time of big bang nucleosynthesis, either at tree level or at one loop. Interesting constraints on this model may come from studies of rare meson decays, flavor observables, and from searches for axion-like particles with mass in the MeV range. We notice however that these small values of $m_a$ are below the sensitivity of BABAR~\cite{Lees:2012iw}, which is the most constraining collider experiment for light pseudo-scalars. It is intriguing that light mediators, with mass around $1$\,--\,$100$ MeV, are advocated by models of self-interacting DM to solve the small scale structures problem of the collisionless DM paradigm \cite{Tulin:2013teo}, although a careful study of the self-interaction potential from the Lagrangian \eqref{eq:lagrangian} is in order to ensure that Coy DM can accommodate the structure anomalies.

\section{Conclusions}\label{sec:concl}

We have shown that a Dirac DM particle interacting with ordinary matter via the exchange of a light pseudo-scalar can accommodate the DAMA data while being compatible with all null direct DM searches. Moreover, it can provide a DM explanation of the GC excess in $\gamma$-rays and achieve the correct relic density. The best fit of both the direct and indirect detection signals is obtained when the pseudo-scalar mediator is much lighter than the DM mass and has universal coupling with heavy quarks, as in hadronic axion models. The leptonic couplings are strongly constrained by precision measurements of the magnetic moment of electron and muon, but they do not enter the analysis and can be safely taken to be zero.

The $99_S\%$ CL compatibility of DAMA with the null searches is determined by the significant enhancement of the coupling to protons with respect to the coupling to neutrons, occurring for natural choices of the pseudo-scalar coupling to quarks. It is intriguing to notice that our results could also be extended to the case of massless mediator since the typical momentum transfer in direct detection is of the order of $m_a$.

Since the phenomenological success of this model relies on the enhancement of the DM-proton coupling respect to the DM-neutron one, as well as on the adopted nuclear form factor, a careful assessment of uncertainties and corrections to these quantities is in order. The model could be tested with measurements of rare meson decays, flavor changing processes, and searches for axion-like particles with mass in the MeV range.

\section{Acknowledgments} 
We thank V.~Cirigliano, N.~Fornengo and F.~Sala for useful discussions. C.A.~and P.P.~acknowledge the support of the ERC project 267117 hosted by UPMC-Paris 6, PI J.~Silk. E.D.N.~acknowledges partial support from the Department of Energy under Award Number DE-SC0009937.

\bibliographystyle{apsrev4-1}
\bibliography{biblio}

\newpage

\appendix
\section{Details on LUX and COUPP likelihood functions}\label{sec:app1}

Here we provide a detailed description of the likelihood functions and the nuisance parameters proper to the LUX and COUPP experiments. Before that, let us notice that details on the treatment of the other experiments included in this letter can be found in \cite{Arina:2011si, Arina:2012dr}, with the following exceptions. We disregard scattering off carbon and chlorine in PICASSO, SIMPLE and COUPP due to the lack of nuclear form factors; however, the most relevant WIMP interaction for these experiments occurs with fluorine, that we consider. For KIMS, WIMP scatterings off cesium and iodine are expected to occur in equal number due to the similar nuclear properties (mass, proton and neutron spin content) of these two elements, see \eg \cite{Toivanen:2009zza}. Due to the lack of a form factor for Cs, we substitute it therefore with the I form factor.

\paragraph*{LUX} The Large Underground Xenon (LUX) experiment consists of a dual-phase xenon detector located at the Sanford Underground Research Facility in the USA. The detector has a fiducial volume of 118 kg and the first science run took place from April to August 2013 for a total of 85.3 live days \cite{Akerib:2013tjd}.

Signals of DM scatterings on nuclei are searched by combining the scintillation light signal (S1) with the secondary ionization signal (S2). In the S1 channel, the detector threshold is set to 2 photoelectrons, which roughly correspond to 3 keVnr (nuclear recoil keV), by using the indicative $L_{\rm eff}$ function in~\cite{szydtalk}. The signal is conservatively set to zero below 3 keVnr, hence the Poisson fluctuations below threshold do not contribute to the estimated signal. The analysis pipeline of LUX is different from the one of XENON100: instead of keeping separated the S1 and S2 signals and use $L_{\rm eff}$, these two quantities are related and modeled with the NEST software~\cite{2011arXiv1106.1613S}. However we will not use this procedure but a simplified approach to specify a likelihood function for LUX.

After cuts, 160 events were observed by the collaboration in a non-blind analysis. Only one event is placed (slightly) below the mean nuclear recoil line extracted from calibration events (see Fig. 4 of~\cite{Akerib:2013tjd}), where a  background of $\bar{B} \pm\sigma_{B} = 0.64\pm 0.16$ events is expected. The likelihood of observing $N=1$ event at fixed signal $S$ and background $B$ is given by the Poisson distribution as 
\begin{multline}
\ln \mathcal{L}_{\rm LUX} (N| S+B)  =   - S + \frac{1}{2} \sigma_B^2 +
\\
\ln\left({\rm e}^{-z^2} \frac{\sigma_B}{\sqrt{2 \pi}} + \frac{1}{2}  \left(S - \sigma_B^2\right)  \left(1 + {\rm erf}(z)\right)\right) ,
\end{multline}
where we have marginalized analytically over the background (as described in~\cite{Arina:2011si}) with $z = (\bar{B} - \sigma_B) / 2 \sigma_B$ and erf the error function. In computing the signal rate we have considered the acceptance as given in the bottom panel of Fig.~1 of~\cite{Akerib:2013tjd} and an additional factor of 1/2 to account for the 50\% nuclear recoil acceptance. With this approximation for the likelihood function there are no nuisance parameters proper to the LUX experiment.

\paragraph*{COUPP} The Chicagoland Observatory for Underground Particle Physics (COUPP) has been operated at the SNOLAB underground laboratory in the USA between September 2010 and August 2011 \cite{Behnke:2012ys}. It consisted of a 4 kg $\rm C F_3 I$ bubble chamber, with fluorine and iodine being sensitive to spin-dependent interactions with protons.

If the energy density injected in the bubble chamber exceeds a certain critical value, a recoiling nucleus traversing the liquid might generate a phase transition \ie a bubble. The detector then operates as a threshold device, controlled by setting the temperature $T$. The relation between the energy threshold $E_{\rm th}(T)$ and the temperature is obtained at a fixed pressure during the calibration process. The observed rate per day per kg of target material is
\begin{eqnarray}\label{eq:rsupheat}
S & = &  \int_{E_{\rm th}(T)}^\infty {\rm d}\ER \, P(\ER, E_{\rm th}(T)) \,  \frac{{\rm d} R}{{\rm d} \ER} \,,
\end{eqnarray}
where $P(\ER, E_{\rm th}(T))$ is a temperature-dependent nucleation efficiency. This is $P(\ER, E_{\rm th}(T)) = \Theta(\ER - E_{\rm th}(T))$ for iodine, while for fluorine it can be parametrized either by
\begin{align}
\label{eq:expcoupp}
P(\ER, E_{\rm th}(T)) &= 1 - \exp \left[ a \left( 1 -\frac{\ER}{E_{\rm th}(T)} \right) \right]
\intertext{or by a step function}
\label{eq:theta}
P(\ER, E_{\rm th}(T)) &= \eta \, \Theta(\ER - E_{\rm th}(T)) \,.
\end{align}
We explore both possibilities. The parameter $a$ defines the steepness of the energy threshold, while $\eta$ has the role of a nucleation efficiency. The values of $a$ and $\eta$ are uncertain and therefore we treat them as nuisance parameters with Gaussian priors centered at $\bar{a} = 0.15 \pm 0.02$ and $\bar{\eta} = 0.49 \pm 0.02$.

The total exposure after cuts is 553 kg-days, subdivided into three run periods, which have a different threshold for the bubble nucleation. The first period is characterized by $N_1=2$ events with an expected background $\bar{B}_1=0.8$ events and has a total exposure of  55.8 kg-days. The second run has $N_2=3$ events, $\bar{B}_2 = 0.7$ events for 70 kg-days, while the third one has $N_3=8$ events, $\bar{B}_3 = 3$ events for 311.7 kg-days. The background comes mainly from neutrons and alpha particles. The exposures take into account the efficiency for single bubble production.

The likelihood is therefore given by the Poisson probability of observing $N$ events in each of the three runs
\begin{equation}
\ln \mathcal{L}_{\rm COUPP}(N|S) =  \sum_{j=1}^3 \ln P(N_j|S+\bar{B}_j) \,.
\end{equation}

By considering the uncertainties on the nucleation parameter (either $a$ or $\eta$) and on the energy thresholds of the three runs, we have four nuisance parameters. For the energy thresholds, we use Gaussian priors with mean values and standard deviations provided in~\cite{Behnke:2012ys}: $\bar{E}^{\rm th}_1 = 7.8$ keVnr, $\sigma_{E_1} = 1.1$ keVnr , $\bar{E}^{\rm th}_2 = 11.0$ keVnr, $\sigma_{E_2} = 1.6$ KeVnr, $\bar{E}^{\rm th}_3 = 15.5$ keV and $\sigma_{E_3} = 2.3$ keVnr.

Finally the likelihood for fluorine in COUPP is given by
\begin{multline}
\ln \mathcal{L}_{\rm COUPP} = \ln\mathcal{L}_{\rm COUPP}(N | S+B)
\\
- \frac{(a -\bar{a})^2}{2 \sigma_{a}^2}
- \sum_{i=1}^3 \frac{(E^{\rm th}_i - \bar{E}^{\rm th}_i)^2}{2 \sigma_{E_i}^2}
\end{multline}
for the nucleation efficiency in \Eq{eq:expcoupp}, and analogous expression for the one in \Eq{eq:theta} with the substitution $a \to \eta$.

We have computed the bounds with both nucleation efficiencies \eqref{eq:expcoupp} and \eqref{eq:theta}. In the region of interest no significant deviation between the two is found since scatterings occur dominantly off iodine, hence we only show the result corresponding to the choice \eqref{eq:expcoupp}.

\section{Details on the DM annihilation cross-section}\label{sec:app2}

The $s$-channel DM annihilation cross section into SM fermions is
\beq
\sigma{(\bar\chi \chi \to \bar f f)}  = N_\text{c} \frac{g^2 g_f^2 g^2_{\rm DM}}{64\pi} \frac{s}{(s - m_a^2)^2} \sqrt{\frac{s - 4 m_f^2}{s - 4 \mDM^2}} \,,
\eeq
where $N_\text{c}$ is the number of colors for the final fermions, and we neglected the pole resonance in the propagator because the mediator is always off-shell for the values of $m_a$ and $\mDM$ needed to explain the DAMA regions.

The DM annihilation cross section into two pseudo-scalars is
\beq
\sigma{(\bar\chi \chi \to \, a a)} = \frac{\gDM^4}{256 \pi} \frac{h(t_0) - h(t_1)}{s (s - 4 \mDM^2)} \,,
\eeq
with
\beq
t_{\genfrac{}{}{0pt}{3}{0}{1}} = - \frac{1}{4} \left( \sqrt{s - 4 \mDM^2} \mp \sqrt{s - 4 m_a^2} \right)^2
\eeq
the integration extrema, and the undefined integral
\begin{multline}
h(t) \equiv 4 \left( \mDM^2 - t \right) + \frac{m_a^4 (u - t)}{\left( \mDM^2 - t \right) \left(\mDM^2 - u \right)}
\\
- \frac{2 m_a^4 + \left( s - 2 m_a^2 \right)^2}{s - 2 m_a^2} \log \! \left( - \frac{\mDM^2 - t}{\mDM^2 - u} \right) ,
\end{multline}
with $u = 2 \mDM^2 + 2 m_a^2 - s - t$.

We compute the thermally averaged annihilation cross section for a non-relativistic DM gas by expanding the cross section in powers of the DM relative velocity $v$, $s \simeq \mDM^2 \left(4 - v^2\right)$, weighting with the appropriate Maxwell-Boltzmann distribution, and then summing over all possible annihilation channels. We obtain
\beq\label{sigmavCoy}
\langle \sigma v \rangle (x) = \sum_f \mathcal A_f + \frac32 \frac{\mathcal B} x + \mathcal O(x^{-2}) \,,
\eeq
where $x \equiv \mDM / T$ with $T$ the temperature of the gas. 
The first coefficient,
\beq\label{eq:Aterm}
\mathcal A_f =  \frac{N_{\rm c}}{8 \pi} \frac{g^2 g_f^2 g^2_{\rm DM} \mDM^2}{(4 \mDM^2 - m_a^2)^2} \sqrt{1- \frac{m^2_f}{\mDM^2}} \,,
\eeq
is the contribution of the s-wave annihilation into SM fermion pairs, while the second coefficient,
\beq\label{eq:Bterm}
\mathcal B = \frac{g_{\rm DM}^4}{96 \pi} \frac{ \mDM^2 (\mDM^2 - m_a^2)^2}{(2 \mDM^2 - m_a^2)^4 }\sqrt{1 - \frac{m^2_a}{\mDM^2}} \,,
\eeq
is given by the annihilation into pseudo-scalar pairs, which occurs in p-wave. The p-wave contribution of the $\bar\chi \chi \to \bar f f$ process is much smaller than the $\bar\chi \chi \to \, a a$ cross section and therefore we neglect it. 

To obtain the value of the thermally averaged cross section at present time, which accounts for the GC $\gamma$-ray excess, we use:
\beq\label{eq:svbest}
2 \langle \sigma v \rangle_{\rm best} = \langle \sigma v \rangle (x_0) \simeq \sum_f \mathcal A_f \,,
\eeq
with $x_0 \gg 1$ the present value of $x$, and $\langle \sigma v \rangle_{\rm best}$ and the adopted value of the DM mass $\mDM^{\rm best}$ taken from Fig. 15 of Ref.~\cite{Daylan:2014rsa}, as explained in the letter. The factor of $2$ in front of $\langle \sigma v \rangle_{\rm best}$ takes into account the fact that $\chi$ is here not self-conjugated, unlike in Ref.~\cite{Daylan:2014rsa}.

\section{Bounds from electron and muon's anomalous magnetic moment}\label{sec:app3}

The presence of a pseudo-scalar mediator coupled to SM fermions may produce detectable effects in various precision measurements, \eg in the  electroweak sector. These observables however usually probe new physics coupled to the electroweak gauge bosons, while the pseudo-scalar state only couples to the SM fermions at tree level and therefore contributes only through two-loop or higher order processes. This, plus the smallness of the couplings favored by DAMA data and the GC excess, makes it easy to exclude any sizable contribution to these observables.

An observable that is able to directly probe new physics coupled to the SM fermions is the anomalous magnetic moment (AMM) of charged leptons.
The AMMs 
of electron and muon are in fact known to a high precision. The experimental values are
\begin{align}
a_e^\text{exp} &= (11596521807.6 \pm 2.7) \times 10^{-13} \, ,
\\
a_\mu^\text{exp} &= (11659209.1 \pm 6.3) \times 10^{-10} \, ,
\end{align}
(as per CODATA recommendations \cite{Mohr:2012tt}, also endorsed by the Particle Data Group \cite{Agashe:2014kda}), while the value predicted by the SM is \cite{KINOSHITA:2014uza, Dorokhov:2014iva}
\begin{align}
a_e^\text{SM} &= (11596521817.8 \pm 7.7) \times 10^{-13} \, ,
\\
a_\mu^\text{SM} &= (11659180.2 \pm 4.9) \times 10^{-10} \, .
\end{align}
The difference between experimental and theoretical values, $\Delta a \equiv a^\text{exp} - a^\text{SM}$, is
\begin{align}
\Delta a_e &= (-10.2 \pm 8.2) \times 10^{-13} \, ,
\\
\label{Delta a_mu}
\Delta a_\mu &= (+28.9 \pm 8.0) \times 10^{-10} \, .
\end{align}
Notice from this last result that there is a $3.6 \sigma$ tension between the measured and theoretical value of the muon's AMM, while for the electron the two are in very good agreement.

\begin{figure}[t!]
\includegraphics[width=0.49\columnwidth]{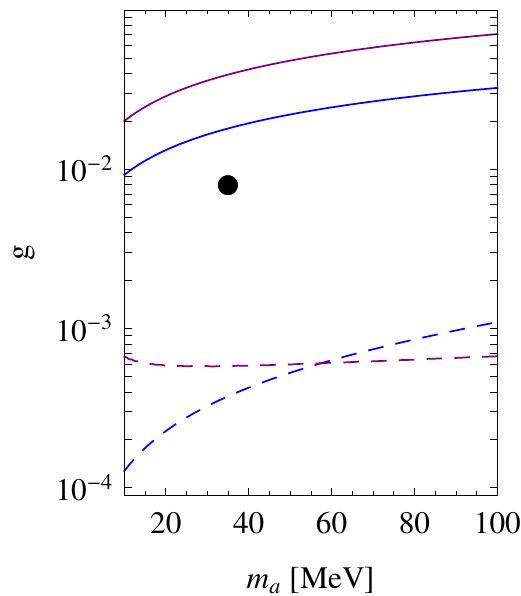}
\includegraphics[width=0.49\columnwidth]{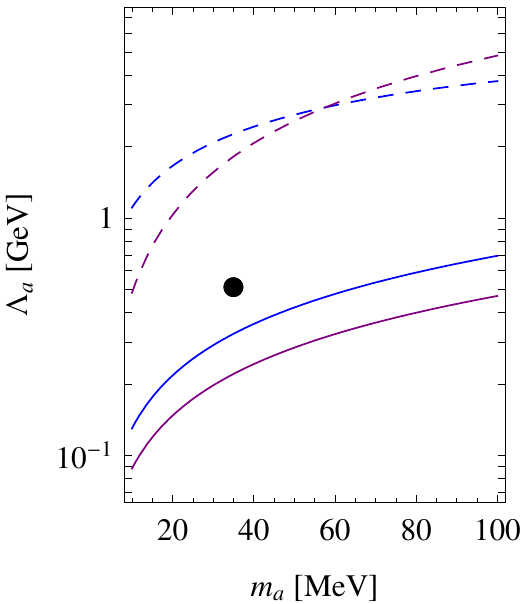}
\caption{\label{fig:AMMu} Upper bound on the pseudo-scalar coupling to SM fermions $g$ (left), and lower bound on $\Lambda_a = m_a / \sqrt{g_{\rm DM} g}$ (right) from electron and muon's AMM. Blue (purple) lines denote bounds from the electron's (muon's) AMM. The dashed lines indicate the two-loop limit in the Universal (\emph{democratic}) case, \ie when the pseudo-scalar couples universally to all SM fermions. The bounds represented by solid lines apply in the leptophobic limit, \ie when the pseudo-scalar couples universally to quarks but does not couple to charged leptons at tree level and therefore the leptons' AMM is generated at three loops. The black dot denotes the best-fitting point to the GC excess for Universal (\emph{democratic}) couplings, as explained in the main text.}
\end{figure}

\begin{figure}[t!]
\includegraphics[width=0.49\columnwidth]{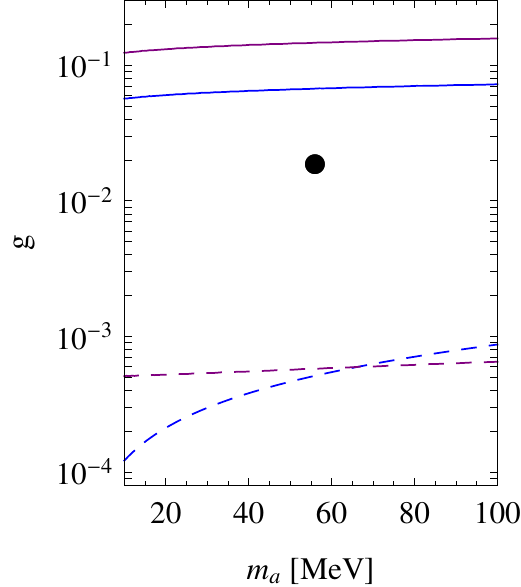}
\includegraphics[width=0.49\columnwidth]{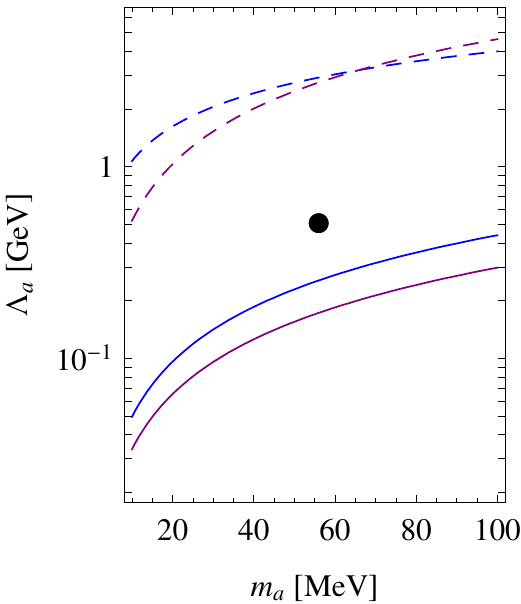}
\caption{\label{fig:AMMh} Same as in Fig.~\ref{fig:AMMu} but for Universal ({\em heavy-flavors}) couplings, \ie when the pseudo-scalar couples universally only to the heavier SM fermions. Dashed lines are for the two-loop limit, while solid lines apply in the leptophobic limit where the first non-zero contribution to the leptons' AMM arises at three loops. The black dot denotes the Universal (\emph{heavy-flavors}) best-fit in parameter space, as explained in the main text.}
\end{figure}

The pseudo-scalar contribution to the AMM of charged leptons has been computed up to two loops. Its phenomenological consequences have been already studied \eg in \cite{Chang:2000ii}, and in \cite{Hektor:2014kga} in the framework of Coy DM, for pseudo-scalar masses above 1 GeV. However we are here interested in masses of the order of tens of MeV. In this regime, the result is dominated by the one-loop contribution which is always negative. Therefore, it is sound to compare the pseudo-scalar contribution to the electron's AMM, $a_e$, with $\Delta a_e$, and requiring that $a_e \leqslant \Delta a_e$. However, the same can not be done with $a_\mu$, since this has opposite sign respect to $\Delta a_\mu$ and therefore the presence of the new particle will only make the deviation of the theoretical result from the experimental measure worse (unless higher loop orders change the sign of $a_\mu$). Therefore we derive a bound from the muon's AMM imposing $a_\mu \leqslant \delta\Delta a_\mu$, where $\delta\Delta a_\mu = 8.0 \times 10^{-10}$ is the error on $\Delta a_\mu$ in Eq.~\eqref{Delta a_mu}. By means of the formulas in \cite{Chang:2000ii}, these bounds can be converted into upper limits on the pseudo-scalar coupling to SM fermions $g$ and lower limits on $\Lambda_a = m_a / \sqrt{g_{\rm DM} g}$ (the parameter of interest for direct DM detection searches), for any given value of the pseudo-scalar mass $m_a$. These limits are shown as dashed curves in Figs.~\ref{fig:AMMu} and \ref{fig:AMMh} (blue for electron, purple for muon), for Universal (\emph{democratic}) and Universal ({\em heavy-flavors}) couplings respectively. The best-fitting points to the GC excess, denoted with black dots, as well as the fit to DAMA data, are excluded in both cases.

Notice that, as said above, these bounds are dominated by the one-loop contribution which only depends on the pseudo-scalar couplings to leptons, but not to quarks. Therefore, the dashed lines in Figs.~\ref{fig:AMMu} and \ref{fig:AMMh} can be intended as bounds on the lepton coupling alone. To avoid these bounds we can assume that the pseudo-scalar state has leptophobic couplings, \ie it doesn't couple to charged leptons. As noted in the main text, the lepton couplings are free parameters that play very little role in fitting the GC excess, and no role whatsoever in direct detection of WIMPs. By setting the couplings of the pseudo-scalar to charged leptons to zero, the one- and two-loop contribution to electron and muon's AMM vanishes, and the first non-zero contribution is expected to arise at three loops. To assess the bound coming from the three-loop-generated AMM, we assume that the most important contribution at this perturbative order is obtained by adding an internal photon line to the two-loop diagrams. Accordingly, we estimate the three-loop $a_\ell$ ($\ell = e, \mu$) to be $\alpha / \pi$ times the two-loop result (when all leptons have been removed from the internal loops), with $\alpha \simeq 1/137$ the electromagnetic fine structure constant. Since we do not know the sign of $a_\ell$, we make the conservative assumption that it has opposite sign respect to $\Delta a_\ell$ for both electron and muon; therefore, we produce bounds on the model parameters by requiring that $a_\ell \leqslant \delta\Delta a_\ell$. The new limits for the leptophobic case are shown as solid lines in Figs.~\ref{fig:AMMu} and \ref{fig:AMMh}. The best-fitting points to the GC excess (as well as the DAMA regions) are now perfectly compatible with the bounds.

\end{document}